\documentclass[12pt]{article}

\usepackage[english]{babel}

\usepackage{latexsym}
\usepackage{amssymb}
\usepackage{epsfig}

\begin{document}

\title{Delocalisation transition in quasi-1D models
with correlated disorder}

\author{ L.~Tessieri${}^{1}$ and F.~M.~Izrailev${}^{2}$ \\
{\it ${}^{1}$ Instituto de F\'{\i}sica y Matem\'{a}ticas} \\
{\it Universidad Michoacana de San Nicol\'{a}s de Hidalgo} \\
{\it Edificio C-3, Ciudad Universitaria} \\
{\it 58060, Morelia, Mich., Mexico} \\
{\it ${}^{2}$ Instituto de F\'{\i}sica, Universidad Aut\'{o}noma de Puebla,} \\
{\it Apdo. Postal J-48, Puebla, Pue. 72570, Mexico}}

\date{4th June 2006}

\maketitle

\begin{abstract}

We introduce a new approach to analyse the global structure of
electronic states in quasi-1D models in terms of the dynamics of a
system of parametric oscillators with time-dependent stochastic
couplings. We thus extend to quasi-1D models the method previously
applied to 1D disordered models. Using this approach, we show that
a ``delocalisation transition'' can occur in quasi-1D models with
weak disorder with long-range correlations.
\end{abstract}

{Pacs numbers:  73.20.Jc, 72.15.Rn, 05.40.-a}

\section{Introduction}

In recent years there has been a steadily increasing interest in
disordered models with long-range correlated disorder. The
interest was initially spurred by the discovery that specific
long-range correlations can produce a kind of ``delocalisation
transition'' even in one-dimensional (1D) models in which the
electronic states are typically exponentially localised~\cite{Mou98}.
The first numerical results were later confirmed by analytical
studies, which identified the relationship existing between
localisation length and pair correlators of the random potential
and showed how to create mobility edges in strictly 1D discrete
models~\cite{Izr99}. These analytical predictions were
experimentally verified by considering the transmission of
microwaves in a single-mode waveguide with a random array of
correlated scatterers~\cite{Kuh00}. Later on, the results obtained
for discrete lattices~\cite{Izr99} were extended to 1D continuous
models~\cite{Izr01} and applied to related problems such as the
propagation of waves in waveguides with random surface scattering,
and to specific quasi-1D models with bulk scattering~\cite{Izr01,Izr03,Izr05}.

One of the main tools for the analysis of such models is based on
the analogy between localisation phenomena in disordered systems
and the dynamics of stochastic oscillators~\cite{Tes00, Tes01}.
This approach was originally applied to strictly 1D models; subsequent
research, however, has begun to explore models of higher dimensionality,
with quasi-1D models providing the first natural extension towards realistic
disordered systems.
A rigorous mathematical treatment of discrete quasi-1D models was
given in~\cite{Roe04}, but the analysis was centred on disorder
without spatial correlations.

This paper serves two main purposes: in the first place, we extend to
systems of higher dimensionality the dynamical approach which was used
successfully for strictly 1D continuous models (see, e.g.,~\cite{IDKT04}
and references therein).
We show that the Lyapunov exponents which govern the exponential divergence
of initially nearby trajectories in a classical systems of stochastic
oscillators are the same exponents which appear in the transfer-matrix
approach for quasi-1D disordered models. This allows us to establish a
rigorous correspondence between the quantum phenomenon of localisation in
quasi-1D disordered models and the orbit instability in classical systems
of parametric oscillators with noisy couplings.

The second objective of this work is to use this analogy to study the
effects of long-range correlations of the disorder on the localisation
of the electronic states in quasi-1D models.
Our main result is that, for weak disorder, specific long-range correlations
can make all Lyapunov exponents vanish (within the second-order
approximation), thereby suppressing orbit instability on the one hand and
producing a ``delocalisation transition'' on the other.

The paper is organised as follows: in Sec.~\ref{corr} we discuss
the correspondence of quasi-1D disordered models with a set of coupled
parametric oscillators. In Sec.~\ref{fokpla} we show how the evolution
of this dynamical system can be analysed. The general results thus obtained
are then applied in Sec.~\ref{longdis} to the specific case in which the
random potential depends only on the longitudinal coordinate.
In Sec.~\ref{deloc} we apply the dynamical approach to the case of a
generic weak disorder. We determine an expression for the sum of the
positive Lyapunov exponents of the quasi-1D model and we use this result
to discuss the delocalisation effects that are produced by specific
long-range correlations of the random potential.
The conclusions are then outlined in Sec.~\ref{conclu}.

\section{Classical representation of the quasi-1D model}
\label{corr}

\subsection{The 1D case}
\label{1d}

Before considering quasi-1D models, we summarise shortly the main results
for the strictly 1D case.
Since this work is focused on quasi-1D models, we shall be brief; the
interested reader can find more details in~\cite{Tes00,Tes01,Tes02}.
The correspondence between Anderson localisation in 1D models with weak
disorder and energetic instability of oscillators with a frequency
perturbed by a noise is a straightforward consequence of the mathematical
analogy between the Schr\"{o}dinger equation
\begin{equation}
-\psi''(x) + U(x) \psi(x) = E \psi(x)
\label{schro1d}
\end{equation}
with positive energy $E$ and the dynamical equation of a stochastic
oscillator
\begin{equation}
\ddot{q}(t) + ( \omega^{2} - U(t) )q(t) = 0
\label{osc1d}
\end{equation}
(here and in the following we will use energy units such that
$\hbar^{2}/2m = 1$).
In fact, Eq.~(\ref{schro1d}) can be easily transformed into Eq.~(\ref{osc1d})
by interpreting the spatial coordinate $x$ as the time $t$ and the
wavefunction amplitude $\psi$ as the coordinate of an oscillator.
In the first equation the function $U$ plays the role of a random potential
while in the second it represents a noise. The noise is white or coloured
depending on whether the disorder is spatially correlated or not.

The mathematical identity of Eqs.~(\ref{schro1d}) and~(\ref{osc1d})
allows one to study the global structure of the quantum eigenstates
of the disordered model~(\ref{schro1d}) by analysing the dynamics of
the corresponding classical oscillator~(\ref{osc1d}).
The dynamical equation~(\ref{osc1d}) gives the time evolution of the
oscillator coordinate $q(t)$ once the initial position $q(0)$ and velocity
$\dot{q}(0)$ have been specified; use of analogous boundary conditions to
solve the Schr\"{o}dinger equation~(\ref{schro1d}) leads to the solution
$\psi(x)$ which is obtained with the standard transfer matrix approach.
Studying the disordered model~(\ref{schro1d}) in terms of the dynamics of
the random oscillator~(\ref{osc1d}), therefore, is equivalent to using
transfer matrix methods.

Comparing the solutions of Eqs.~(\ref{schro1d}) and~(\ref{osc1d}) one
finds that spatially extended states correspond to bounded oscillator
orbits, while localised states have their counterpart in unbounded
trajectories.
As a consequence, the phenomenon of Anderson localisation corresponds
to energetic instability of the parametric oscillator, with the inverse
localisation length being equal to the Lyapunov exponent of the stochastic
oscillator, i.e., the rate of exponential divergence for initially nearby
orbits.

The case of weak noise/disorder can be studied using perturbative techniques
which were originally devised for the study of stochastic systems. These
methods allow one to obtain the rate of energy growth for the
oscillator~(\ref{osc1d}) and therefore the inverse localisation length for
the disordered model~(\ref{schro1d})~\cite{Tes01, Tes02}.
The second-order expression for the inverse localisation length is
\begin{equation}
\lambda = \frac{1}{4 \omega^{2}} \int_{0}^{\infty}
\langle U(t) U(t+\tau) \rangle \cos \left( 2 \omega \tau \right) d\tau
\label{invloc1d}
\end{equation}
which shows that the Lyapunov exponent is proportional to the power
spectrum of the disorder, i.e., to the cosine Fourier transform of the
two-point correlator of the random potential (here and in what follows
we use the symbol $\langle \ldots \rangle$ to denote the average over
different disorder realisations).
This entails that the energetic instability of the oscillator~(\ref{osc1d})
can be suppressed if the unperturbed oscillator frequency, multiplied by a
factor two, lies in a frequency interval where the power spectrum of the
disorder vanishes.
Correspondingly, delocalised states arise for the energy values for which
the inverse localisation length~(\ref{invloc1d}) vanishes.

\subsection{The quasi-1D model}

We analyse the phenomenon of electronic localisation in a
semi-infinite strip. The present method can be applied to bars
as well; we focus on strips to avoid unnecessary complications in
the mathematical formulae.
We consider the strip $D$ in the $x-y$ plane
\begin{equation}
D = \{ (x,y): 0 \leq x; \; 0 \leq y \leq L \} .
\label{domain}
\end{equation}
The Schr\"{o}dinger equation for a quantum particle (``electron'') in
the strip is
\begin{equation}
- \left( \frac{\partial^{2}}{\partial x^{2}}
+ \frac{\partial^{2}}{\partial y^{2}}\right) \psi(x,y)
+ \varepsilon U(x,y) \psi(x,y) =
E \psi(x,y) .
\label{schroe1}
\end{equation}
The function $U(x,y)$ is the random potential felt by the electrons; the
potential can exhibit spatial correlations. The dimensionless parameter
$\varepsilon$ is introduced to keep track of the order of perturbative
expansions and can be set equal to $1$ in the final formulae.
We will focus on the case of weak disorder, i.e., on the case $\varepsilon
\ll 1$.
For the model to be completely defined, one must provide the statistical
properties of the random potential; for weak disorder, it is enough to
specify the first two moments of $U(x,y)$. We will assume that the potential
has zero average, $\langle U(x,y) \rangle = 0$, and that the two-point
correlator is a known function
\begin{equation}
\langle U(x_{1},y_{1}) U(x_{2},y_{2}) \rangle = \sigma^{2}
\chi(x_{1}-x_{2}, y_{1}-y_{2})
\label{bincor}
\end{equation}
where $\sigma^{2}$ represents the variance of the potential
\begin{displaymath}
\sigma^{2} = \langle U(x,y) U(x,y) \rangle
\end{displaymath}
provided that it is finite; when the variance diverges, like in the case
of white noise, $\sigma^{2}$ must be interpreted as a parameter which
measures the strength of the disorder.
We will consider random potentials which are translationally invariant in
the mean; this is why we assume that the binary correlator~(\ref{bincor})
depends only on the difference of the coordinates of the two points
$(x_{1},y_{1})$ and $(x_{2},y_{2})$.

\subsection{Representation of the Schr\"{o}dinger equation in the partially
reciprocal space}

If one wishes to extend to quasi-1D models the dynamical approach described
in subsection~\ref{1d} for the 1D case, it is quite natural to identify the
longitudinal coordinate $x$ with the time $t$ of the corresponding dynamical
system. One can then Fourier-transform the Schr\"{o}dinger
equation~(\ref{schroe1}) in the transversal direction $y$.
In the following we consider the real solutions of the stationary
Schr\"{o}dinger equation~(\ref{schroe1}); this choice is not restrictive
because one can find a complete set of real eigenfunctions of the
Hamiltonian.
It is possible to expand any real function of $y$ defined over the interval
$[0:L]$ in a cosine basis. In this way one avoids dealing with complex
Fourier components, which makes easier the successive identification of the
Fourier components with oscillator coordinates discussed below.
Expanding the wavefunction and the potential one obtains
\begin{equation}
\begin{array}{ccl}
\psi(x,y) & = & \displaystyle
\tilde{\psi}_{0}(x) + 2 \sum_{n=1}^{\infty} \tilde{\psi}_{n}(x)
\cos \left(\frac{\pi n y}{L}\right) = \sum_{n=-\infty}^{\infty}
\tilde{\psi}_{n}(x) \cos \left(\frac{\pi n y}{L}\right) \\
U(x,y) & = & \displaystyle
\tilde{U}_{0}(x) + 2 \sum_{n=1}^{\infty} \tilde{U}_{n}(x)
\cos \left(\frac{\pi n y}{L}\right) = \sum_{n=-\infty}^{\infty}
\tilde{U}_{n}(x) \cos \left(\frac{\pi n y}{L}\right)
\end{array}
\label{invft}
\end{equation}
where the Fourier components of the expansion are defined by
\begin{equation}
\tilde{\psi}_{n}(x) = \frac{1}{L}
\int_{0}^{L} dy \; \psi(x,y) \cos \left(\frac{\pi n y}{L}\right)
\label{ft}
\end{equation}
and
\begin{equation}
\tilde{U}_{n}(x) = \frac{1}{L} \int_{0}^{L} dy \; U(x,y)
\cos \left(\frac{\pi n y}{L}\right) .
\label{uft}
\end{equation}
In the following we will often refer to the Fourier components of the
wavefunctions as Fourier modes or channels.
Note that the Fourier components with opposite values of the index $n$
are identical: $\tilde{\psi}_{n}(x) = \tilde{\psi}_{-n}(x)$ and
$\tilde{U}_{n}(x) = \tilde{U}_{-n}(x)$.
The functions defined by the Fourier cosine expansion~(\ref{invft})
coincide with the original functions in the domain~(\ref{domain});
outside of this domain, they are symmetric and $2L$-periodic functions of
the transversal coordinate $y$, i.e.,
\begin{eqnarray*}
\psi(x,y) = \psi(x,-y) & \mbox{ and} &  \psi(x,y) = \psi(x,y+2L) .
\end{eqnarray*}
The same relations also hold for $U(x,y)$.
Expanding in a cosine series, therefore, is equivalent to consider
$y$-symmetric functions $\psi(x,y)$ and $U(x,y)$ on the doubled strip
$[0:\infty] \times [-L:L]$ and impose periodic boundary conditions along
the transversal direction (i.e., to roll the doubled strip into a cylinder).
Boundary conditions can be chosen with considerable latitude because they
do not influence the structure of the electronic state if the strip is
sufficiently wide.
By taking the Fourier cosine transform, one can write Eq.~(\ref{schroe1})
in the form
\begin{equation}
\frac{\partial^{2} \tilde{\psi}_{n}}{\partial x^{2}}(x) +
\left[ E - \left( \frac{\pi n}{L} \right)^{2} \right]
\tilde{\psi}_{n}(x) = \sum_{k = -\infty}^{\infty} \varepsilon
\tilde{U}_{n-k}(x) \tilde{\psi}_{k}(x) .
\label{schroe2}
\end{equation}

\subsection{Elliptic and hyperbolic Fourier components}

Before giving a dynamical interpretation for Eq.~(\ref{schroe2}),
it is useful to introduce a distinction between ``elliptic'' and
``hyperbolic'' Fourier components of the wavefunction.
Adopting the terminology of~\cite{Roe04}, we define the Fourier component
$\tilde{\psi}_{n}(y)$ with wave number $n$ to be
\begin{displaymath}
\begin{array}{lcr}
\mbox{elliptic} & \mbox{if} & \displaystyle
E - \left( \frac{\pi n}{L} \right)^{2} > 0 \\
\mbox{hyperbolic} & \mbox{if} & \displaystyle
E - \left( \frac{\pi n}{L} \right)^{2} < 0
\end{array}
\end{displaymath}
(we ignore the marginal case of the ``parabolic'' components with
$E = \left( \pi n/L \right)^{2}$).

The crucial difference between elliptic and hyperbolic components is that the
latter decay exponentially in the longitudinal direction: from a physical
point of view, they are evanescent modes (or closed channels). Hence they
can be neglected for large values of the longitudinal coordinate $x$.
The irrelevance of the hyperbolic modes can be justified also with a
different argument. As can be seen from the Fourier expansions~(\ref{invft}),
the Fourier components with large values of $n$ describe the behaviour of
the wavefunction on small spatial scales
\begin{displaymath}
\delta y \sim \frac{L}{n}
\end{displaymath}
in the transversal direction.
If we suppose that our continuous model has an underlying lattice structure,
however, we need not consider the behaviour of the wavefunction over
spatial scales $\delta y \lesssim a$, where $a$ is the lattice constant.
This implies that, if the short-scale spatial structure of the
wavefunction is neglected, the only Fourier components that must
be taken into account are the ones with wavenumber
\begin{equation}
n < \overline{n}  \sim \frac{L}{a} .
\label{cutoff}
\end{equation}
Now, let us restrict our analysis of localisation to the case of high-energy
electrons, where high-energy means
\begin{equation}
E > \left( \frac{\pi \overline{n}}{L} \right)^{2} \sim
\left( \frac{\pi}{a} \right)^{2} .
\label{highen}
\end{equation}
In this case, one can see that, due to the cutoff condition~(\ref{cutoff}),
the only Fourier components which are relevant for the analysis are
the elliptic ones.
Hence, we need not bother with hyperbolic components if we restrict our
attention to the regime of sufficiently high energies.
This back-of-the-envelope criterion suggests that our approximation
may fail when the electron energy does not satisfy condition~(\ref{highen}).

After eliminating the hyperbolic components, the Schr\"{o}dinger
equation~(\ref{schroe2}) takes the form
\begin{equation}
\frac{\partial^{2} \tilde{\psi}_{n}}{\partial x^{2}}(x) +
\left[ E - \left( \frac{\pi n}{L} \right)^{2} \right]
\tilde{\psi}_{n}(x) = \sum_{k = -N}^{N} \varepsilon \tilde{U}_{n-k}(x)
\tilde{\psi}_{k}(x)
\label{newschroe}
\end{equation}
where the indices $n$ and $k$ are restricted to the elliptic modes (or
conducting channels), i.e., the indices take the values $-N, -N+1, \ldots,
N-1,N$ with $N$ being the integer part $[\ldots]$ of the ratio
$L\sqrt{E}/\pi$,
\begin{displaymath}
N = \left[ \frac{L \sqrt{E}}{\pi} \right] .
\end{displaymath}

\subsection{Dynamical equations for the quasi-1D model}

In order to give a dynamical interpretation of the Schr\"{o}dinger
equation~(\ref{newschroe}) we replace the longitudinal variable $x$ with
a time variable $t$. In addition, we define the frequencies
\begin{equation}
\omega_{n} = \sqrt{ E - \left( \frac{\pi n}{L} \right)^{2}}
\label{freque}
\end{equation}
and we introduce the notation
\begin{displaymath}
q_{n}(t) = \tilde{\psi}_{n}(t)
\end{displaymath}
for the Fourier components of the wavefunction.
This allows us to write Eq.~(\ref{newschroe}) in the form
\begin{equation}
\ddot{q}_{n}(t) + \omega_{n}^{2} q_{n}(t) = \varepsilon \sum_{k=-N}^{N}
\tilde{U}_{n-k}(t) q_{k}(t)
\label{dyneq}
\end{equation}
which is naturally interpreted as the dynamical equation of a system of
classical parametric oscillators with time-dependent stochastic couplings.
This correspondence allows one to analyse the structure of the electronic
states of the disordered model~(\ref{schroe1}) in terms of the dynamics of
the system~(\ref{dyneq}). In fact, the spatial behaviour of the
$\tilde{\psi}_{n}(x)$ Fourier component along the longitudinal direction
is determined by the time evolution of the coordinate $q_{n}(t)$ of the
corresponding oscillator.

The system of $2N+1$ second-order differential equations~(\ref{dyneq})
can be transformed into a system of $4N+2$ first-order Hamiltonian
equations by introducing the momenta $p_{n} = \dot{q}_{n}$ and the
Hamiltonian
\begin{equation}
H(p,q) = \sum_{n=-N}^{N} \left( \frac{p_{n}^{2}}{2} + \frac{\omega_{n}^{2}}{2}
q_{n}^{2} \right) - \varepsilon \sum_{n=-N}^{N} \sum_{k=-N}^{N}
\tilde{U}_{n-k}(t) q_{n}q_{k} .
\label{ham1}
\end{equation}
Then one can cast the dynamical system~(\ref{dyneq}) in the form
\begin{equation}
\begin{array}{ccl}
\dot{p}_{n} & = & \displaystyle - \frac{\partial H}{\partial q_{n}} =
- \omega_{n}^{2} q_{n} + \varepsilon \sum_{k=-N}^{N} \tilde{U}_{n-k}(t)
q_{k} \\
\dot{q}_{n} & = & \displaystyle
\frac{\partial H}{\partial p_{n}} = p_{n} .
\end{array}
\label{hameq}
\end{equation}

Note that these Hamiltonian equations describe a system of $2N+1$ oscillators.
However, not all oscillators are independent, because the symmetry of the
Fourier components $\tilde{\psi}_{n}(x) = \tilde{\psi}_{-n}(x)$ implies
that the Hamiltonian system is subject to the constraints $q_{n} = q_{-n}$
which reduce to $N+1$ the number of degrees of freedom of the system.
Making use of the constraints, the dynamical equations~(\ref{hameq}) can
be written in the form
\begin{equation}
\begin{array}{ccl}
\dot{p}_{n} & = & \displaystyle
- \omega_{n}^{2} q_{n} + \sum_{k=0}^{N} \varepsilon W_{nk}(t) q_{k} \\
\dot{q}_{n} & = & \displaystyle = p_{n}
\end{array}
\label{dynsys}
\end{equation}
where $n=0,1,\ldots,N$ and only independent oscillators are involved.
Note that in Eq.~(\ref{dynsys}) we have introduced the short-hand
notation
\begin{equation}
W_{n,k}(t) = \left[ \tilde{U}_{n+k}(t) + \tilde{U}_{n-k}(t) \right]
\left( 1 - \frac{1}{2} \delta_{k 0} \right) .
\label{wmatrix}
\end{equation}

A comment is in order here. A mapping of a quasi-1D disordered model of
the form~(\ref{schroe1}) unto a Hamiltonian system was made long ago
by Hansel and Luciani~\cite{Han87, Han89}. In spite of several similarities,
their approach differs from the present one in many aspects.
In the first place, in~\cite{Han87, Han89} the quasi-1D model is morphed
into a Hamiltonian system by {\em discretising} rather than {\em
Fourier-transforming} the system in the transversal directions. In our
approach, no discretisation is required. This implies that the channels
Hansel and Luciani deal with are 1D chains coupled linearly to each other,
whereas in our scheme the channels are the Fourier modes of the wavefunction.
Because of this conceptual difference, our channels are coupled only if the
random potential is present, contrary to what happens in the Hansel and
Luciani work. Similarly, no distinction between elliptic and hyperbolic
channels arise in that case.
In the second place, the Hamiltonian model of Hansel and Luciani has no
constraints, while the system~(\ref{ham1}) is subject to the constraints
$q_{n} = q_{-n}$. This entails that the coupling term~(\ref{wmatrix})
is not fully symmetric in the two indices $n$ and $k$, unlike the coupling
matrix considered in~\cite{Han87, Han89}.

\subsection{Wavefunction localisation versus oscillator instability}
\label{analogy}

The mapping of Eq.~(\ref{newschroe}) unto Eq.~(\ref{dynsys})
makes possible to study the behaviour of the wavefunctions of the disordered
model~(\ref{schroe1}) in terms of the time evolution of the dynamical
system~(\ref{dynsys}). From a physical point of view, the mathematical
identity of the two problems translates into a correspondence between
the phenomenon of Anderson localisation in quasi-1D disordered bars and the
energetic instability of the system of random oscillators.
The connection between the two phenomena can be understood and made
quantitative by observing that the crucial properties of both classes of
systems are defined in terms of their Lyapunov exponents. In the case of
quasi-1D systems, Lyapunov exponents arise within the framework of the
transfer matrix approach (see, e.g.,~\cite{Mac03} and references therein).
The method divides the strip (or the bar) in layers; a transfer matrix
is an operator that relates the values of the wavefunction and its
derivative on one layer with the corresponding values on the contiguous
layer. The Schr\"{o}dinger equation is considered as an initial value
problem and its solution is obtained in terms of a product of transfer
matrices (in discrete models) or of spatial-ordered exponentials (for
continuous models).
Making use of Oseledec's theorem~\cite{Ose68}, one can then define the
Lyapunov exponents for the quasi-1D model in terms of the eigenvalues of
the asymptotic product of transfer matrices. The localisation length
is equal to the smallest positive Lyapunov exponent.

For dynamical systems, the Lyapunov characteristic exponents can be
introduced by considering the exponential divergence of initially nearby
orbits. For maps, this leads to the computation of the eigenvalues of
limit product of matrices~(see, e.g.,~\cite{Ott93}) which give the
evolution of vectors in tangent space and are the counterpart of the
transfer matrices for the quasi-1D models.
To sum up, Lyapunov exponents can be defined both for quasi-1D models and
for dynamical systems; in particular, due to the mathematical identity of
Eqs.~(\ref{newschroe}) and~(\ref{dynsys}), the (non-negative parts of the)
two Lyapunov spectra are the same for both systems.

From this point of view, quasi-1D models are similar to 1D systems of the
type described by Eq.~(\ref{schro1d}), whose single Lyapunov exponent (i.e.,
the inverse of the localisation length) is identical to the Lyapunov exponent
of the dynamical counterpart~(\ref{osc1d}) (i.e., the rate of exponential
orbit divergence).
However, in the case of quasi-1D systems, there is an important difference
between the disordered model~(\ref{schroe1}) and its dynamical
analogue~(\ref{dynsys}): for the former, the most important Lyapunov
exponent is the {\em smallest} one, which defines the inverse localisation
length, while for the latter the dynamics is dominated by the {\em largest}
Lyapunov exponent, which gives the rate of exponential divergence along
almost every direction of the phase space of the system.
Therefore, in order to extract information on the spatial behaviour of
the electronic states of model~(\ref{schroe1}), a thorough analysis of the
time evolution of the dynamical system~(\ref{dynsys}) is needed.
In particular, one cannot restrict the dynamical analysis to the
determination of the largest Lyapunov exponent.
Having thus highlighted the links between the quantum phenomenon of
localisation and the dynamics of a system of parametric oscillators,
we now turn our attention to the study of the latter problem.

\section{Dynamics of the parametric oscillators}
\label{fokpla}

To analyse the dynamics of a system of oscillators, it is convenient to
perform a canonical transformation and switch from the Cartesian coordinates
$(q_{n},p_{n})$ to the action-angle variables $(J_{n},\theta_{n})$ defined
by the relations
\begin{displaymath}
\begin{array}{ccl}
q_{n} & = & \displaystyle \sqrt{\frac{2J_{n}}{\omega_{n}}} \sin \theta_{n} \\
p_{n} & = & \displaystyle \sqrt{2 \omega_{n} J_{n}} \cos \theta_{n} .
\end{array}
\end{displaymath}
In terms of the new variables the Hamiltonian~(\ref{ham1}) takes the form
\begin{displaymath}
H = \sum_{n=-N}^{N} \omega_{n} J_{n} - \sum_{n=-N}^{N} \sum_{k=-N}^{N}
\varepsilon \tilde{U}_{n-k}(t) \sqrt{\frac{J_{n}J_{k}}{\omega_{n} \omega_{k}}}
\sin \theta_{n} \sin \theta_{k}
\end{displaymath}
and the dynamical equations~(\ref{dynsys}) become
\begin{equation}
\begin{array}{ccl}
\dot{J}_{n} & = & \displaystyle
\sum_{k=0}^{N} 2 \varepsilon W_{n,k}(t)
\sqrt{\frac{J_{n}J_{k}}{\omega_{n} \omega_{k}}}
\sin \theta_{k} \cos \theta_{n} \\
\dot{\theta}_{n} & = & \displaystyle
\omega_{n} - \sum_{k=0}^{N} \varepsilon W_{n,k}(t)
\sqrt{\frac{J_{k}}{J_{n} \omega_{n} \omega_{k}}}
\sin \theta_{k} \sin \theta_{n} .
\end{array}
\label{langevin}
\end{equation}

Following the general method described by Van Kampen~\cite{Van92}, one
can replace the system of Langevin equations~(\ref{langevin}) with a
deterministic Fokker-Planck equation for the probability distribution
of the stochastic variables $(J_{n},\theta_{n})$.
In fact, the dynamical equations~(\ref{langevin}) are stochastic equations
of the form
\begin{equation}
\dot{u} = F^{(0)}(u) + \varepsilon F^{(1)}(u,t)
\label{stoceq}
\end{equation}
where $F^{(0)}(u)$ is a sure function of the vector $u$ and
$\varepsilon F^{(1)}(u,t)$ is a stochastic correction to the
deterministic term. Following Van Kampen~\cite{Van92}, one can associate
to the stochastic differential equation~(\ref{stoceq}) an ordinary
differential equation for the function $P(u,t)$ which represents the
probability distribution for the random variable $u$.
The associated differential equation has the form
\begin{equation}
\begin{array}{l}
\displaystyle
\frac{\partial P}{\partial t} (u,t) =
- \sum_{i} \frac{\partial}{\partial u_{i}} \left[ F_{i}^{(0)}(u)
P(u,t) \right] \\
\displaystyle +
\varepsilon^{2} \sum_{ij} \frac{\partial}{\partial u_{i}}
\int_{0}^{\infty} d\tau \langle F_{i}^{(1)}(u,t)
\frac{d\left( u^{-\tau}\right)}{d \left( u \right)}
\frac{\partial}{\partial u_{j}^{-\tau}} F_{j}^{(1)}
\left( u^{-\tau},t-\tau \right) \rangle
\frac{d \left( u \right)}{d \left( u^{-\tau} \right)} P(u,t)
+ o(\varepsilon^{2}),
\end{array}
\label{genfp1}
\end{equation}
where $u^{t}$ is the solution at time $t$ of the unperturbed differential
equation $\dot{u} = F^{(0)}(u)$ with the initial condition $u(0) = u$,
and $d \left( u^{-\tau} \right)/d \left( u \right)$ is the Jacobian
of the transformation $u \rightarrow u^{-\tau}$.
Note that, within the second-order approximation, i.e., neglecting
terms of order $o(\varepsilon^{2})$, only first- and second-order
derivative appear in the right hand side (r.h.s.) of Eq.~(\ref{genfp1}),
which therefore has the form of a Fokker-Planck equation.

In the present case, the vectors $u$, $F^{(0)}(u)$, and $F^{(1)}(u,t)$
have the $n$-th bi-component equal to
\begin{displaymath}
\begin{array}{ccc}
u_{n} = \left( \begin{array}{c} J_{n} \\
                                   \theta_{n}
                   \end{array} \right), &
F_{n}^{(0)} = \left( \begin{array}{c} 0 \\
                                   \omega_{\bf n}
                   \end{array} \right), &
F_{n}^{(1)} = \left( \begin{array}{c} F_{n,J}^{(1)}(J,\theta) \\
                                      F_{n,\theta}^{(1)}(J,\theta)
                   \end{array} \right)
\end{array}
\end{displaymath}
where we have introduced the symbols
\begin{displaymath}
\begin{array}{ccc}
\displaystyle
F_{n,J}^{(1)}(J,\theta) & = & \displaystyle
\sum_{k=0}^{N} 2 W_{n,k}(t) \sqrt{\frac{J_{n}J_{k}}{\omega_{n} \omega_{k}}}
\sin \theta_{k} \cos \theta_{n} \\
\displaystyle
F_{n,\theta}^{(1)}(J,\theta) & = & \displaystyle
- \sum_{k=0}^{N} W_{n,k}(t) \sqrt{\frac{J_{k}}{J_{n} \omega_{n} \omega_{k}}}
\sin \theta_{k} \sin \theta_{n}
\end{array} .
\end{displaymath}
The unperturbed flow $u \rightarrow u^{t}$ in this specific case has the
simple form
\begin{displaymath}
\begin{array}{ccccc}
u_{n}^{t} & = & \left( \begin{array}{c} J_{n}^{t} \\
                                        \theta_{n}^{t}
                       \end{array} \right) & = &
\left( \begin{array}{c} J_{n} \\
                        \omega_{n}t + \theta_{n}
       \end{array} \right)
\end{array}
\end{displaymath}
and as a consequence one has
\begin{displaymath}
\begin{array}{ccc}
\displaystyle
\frac{d \left( u^{-\tau} \right)}{d \left( u \right)} =
\frac{d \left( u \right)}{d \left( u^{-\tau} \right)} = 1 &
\mbox{and } & \displaystyle
\frac{\partial}{\partial u_{i}^{-\tau}} = \frac{\partial}{\partial u_{i}}
\end{array} .
\end{displaymath}
Using these results, the general Fokker-Planck equation~(\ref{genfp1})
takes the specific form
\begin{equation}
\begin{array}{l}
\displaystyle
\frac{\partial P}{\partial t} (J, \theta, t) = - \sum_{n=0}^{N}
\omega_{n} \frac{\partial P}{\partial \theta_{n}} (J, \theta, t) \\
\displaystyle
+ \varepsilon^{2}
\sum_{n=0}^{N} \sum_{k=0}^{N} \frac{\partial}{\partial J_{n}}
\int_{0}^{\infty} d\tau \langle F_{n,J}^{(1)}(J,\theta,t)
\frac{\partial}{\partial J_{k}}
F_{k,J}^{(1)}(J,\theta - \omega \tau, t - \tau) \rangle P(J,\theta,t) \\
\displaystyle
+ \varepsilon^{2}
\sum_{n=0}^{N} \sum_{k=0}^{N} \frac{\partial}{\partial J_{n}}
\int_{0}^{\infty} d\tau \langle F_{n,J}^{(1)}(J,\theta,t)
\frac{\partial}{\partial \theta_{k}}
F_{k,\theta}^{(1)}(J,\theta - \omega \tau, t - \tau) \rangle
P(J,\theta,t) \\
\displaystyle
+ \varepsilon^{2}
\sum_{n=0}^{N} \sum_{k=0}^{N} \frac{\partial}{\partial \theta_{n}}
\int_{0}^{\infty} d\tau \langle F_{n,\theta}^{(1)}(J,\theta,t)
\frac{\partial}{\partial J_{k}}
F_{{k},J}^{(1)}(J,\theta - \omega \tau, t - \tau) \rangle P(J,\theta,t) \\
\displaystyle
+ \varepsilon^{2}
\sum_{n=0}^{N} \sum_{k=0}^{N} \frac{\partial}{\partial \theta_{n}}
\int_{0}^{\infty} d\tau \langle F_{n,\theta}^{(1)}(J,\theta,t)
\frac{\partial}{\partial \theta_{k}}
F_{k,\theta}^{(1)}(J,\theta - \omega \tau, t - \tau) \rangle
P(J,\theta,t)
\end{array}
\label{fp0}
\end{equation}
Rather than trying to obtain the general solution of Eq.~(\ref{fp0}) -an
exceedingly difficult task-, we aim at deriving the reduced probability
distribution for the action variables
\begin{displaymath}
P(J,t) = \int \prod_{n} d \theta_{n} P(J,\theta,t) .
\end{displaymath}
Since the frequencies $\omega_{n}$ are different from zero, a glance
at the dynamical equations~(\ref{langevin}) shows that the angular
variables are ``fast'' in comparison with the ``slow'' action variables.
We can therefore assume that, after a sufficiently long time, the angle
variables become uncorrelated random variables with a uniform distribution
in the $[0:2 \pi]$ interval, i.e., that for large times the distribution
$P(J,\theta,t)$ reduces to the factorised form
\begin{equation}
P(J,\theta,t) \simeq \frac{1}{(2 \pi)^{N+1}} P(J,t) .
\label{flatangle}
\end{equation}
Substituting the distribution~(\ref{flatangle}) in the Fokker-Planck
equation~(\ref{fp0}) and integrating over the angular variables, one
obtains, after some algebra, the reduced Fokker-Planck equation
\begin{equation}
\frac{\partial P}{\partial t}(J,t) = \sum_{n=0}^{N}
\sum_{k=0}^{N} \frac{1}{2} \frac{\partial}{\partial J_{n}} \left[
D_{nk}(J) \frac{\partial P}{\partial J_{k}}(J,t) \right]
\label{fp1}
\end{equation}
where the diffusion matrix $D_{nk}(J)$ has diagonal elements defined as
\begin{equation}
\begin{array}{ccl}
D_{nn}(J) & = & \displaystyle
\left( \frac{\varepsilon J_{n}}{\omega_{n}} \right)^{2}
\int_{0}^{\infty} d \tau \;
\langle W_{nn}(t) W_{nn}(t-\tau) \rangle \cos 2 \omega_{n} \tau \\
& + & \displaystyle
\sum_{k \neq n} \frac{\varepsilon^{2} J_{n}J_{k}}{\omega_{n} \omega_{k}}
\int_{0}^{\infty} d \tau \; \langle W_{nk}(t) W_{nk}(t-\tau) \rangle
\left[ \cos \left( \omega_{n} + \omega_{k} \right) \tau +
\cos \left( \omega_{n} - \omega_{k} \right) \tau \right]
\end{array}
\label{dmatrix1}
\end{equation}
and off-diagonal elements ($n \neq k$) equal to
\begin{equation}
D_{nk}(J) = \frac{\varepsilon^{2} J_{n}J_{k}}{\omega_{n} \omega_{k}}
\int_{0}^{\infty} d \tau \; \langle W_{nk}(t) W_{kn}(t-\tau) \rangle
\left[ \cos \left( \omega_{n} + \omega_{k} \right) \tau -
\cos \left( \omega_{n} - \omega_{k} \right) \tau \right] .
\label{dmatrix2}
\end{equation}

The Fokker-Planck equation~(\ref{fp1}) is a rather general result, because
it has been obtained for a generic random potential, with the only
assumption that the disorder should be weak and translationally invariant
in the mean.
At first sight, one could find surprising that a localisation phenomenon
can be described with a diffusion equation. The apparent paradox can be
explained by considering that the localised Fourier components of the
electronic wavefunction have oscillators with unbounded orbits as
dynamical counterpart. This equivalence provides an intuitive reason
why electronic localisation can be analysed with a diffusion equation
for the action variables.

We note that only second-order derivatives appear in the diffusion
equation~(\ref{fp1}), and that the coefficients of its r.h.s. are
completely determined by the two-point correlator of the random
potential. Both facts are consequences of the decision to study the
problem within the framework of the second-order approximation.
Pushing the perturbative approach to higher orders, in fact, would
make higher-order derivatives appear in the diffusion equation, together
with coefficients which would depend on higher-order moments of the
potential. It may be appropriate to stress that, if in the present work
the statistical properties of disorder are defined only up to the binary
correlator, this is not due to any Gaussian assumption but is only a
consequence of the adopted second-order perturbative scheme, which makes
unnecessary to specify the statistical features of the random potential
beyond the second moments.

Using the Fokker-Planck equation~(\ref{fp1}) for the reduced probability
density $P(J,t)$ allows one to obtain the dynamical equation for the
averaged action variables
\begin{displaymath}
\overline{J_{n}} = \int dJ_{0} \ldots dJ_{N} \; J_{n} P(J_{0},\ldots,J_{N},t) .
\end{displaymath}
In fact, differentiating with respect to time both members of the previous
equation and using Eq.~(\ref{fp1}) to express the time derivative of
$P(J,t)$, one arrives at the matrix equation
\begin{equation}
\frac{d}{dt} \left( \begin{array}{c} \overline{J_{0}} \\
                                     \vdots \\
                                     \overline{J_{N}} \\
                    \end{array} \right) =  {\bf M}
\left( \begin{array}{c} \overline{J_{0}} \\
                                     \vdots \\
                                     \overline{J_{N}} \\
       \end{array} \right)
\label{avj}
\end{equation}
where ${\bf M}$ is a $(N+1)\times(N+1)$ matrix with diagonal elements
\begin{displaymath}
\begin{array}{ccl}
{\bf M}_{nn} & = & \displaystyle
\frac{\varepsilon^{2}}{\omega_{n}^{2}} \int_{0}^{\infty} d \tau \;
\langle W_{nn}(t) W_{nn}(t-\tau) \rangle \cos 2 \omega_{n} \tau \\
& + & \displaystyle
\sum_{k \neq n} \frac{\varepsilon^{2}}{2 \omega_{n} \omega_{k}}
\int_{0}^{\infty} d \tau \; \langle W_{nk}(t) W_{kn}(t-\tau) \rangle
\left[ \cos \left( \omega_{n} + \omega_{k} \right) \tau -
\cos \left( \omega_{n} - \omega_{k} \right) \tau \right]
\end{array}
\end{displaymath}
and off-diagonal elements ($n \neq k$)
\begin{displaymath}
{\bf M}_{nk} = \frac{\varepsilon^{2}}{2 \omega_{n} \omega_{k}}
\int_{0}^{\infty} d \tau \; \langle W_{nk}(t) W_{nk}(t-\tau) \rangle
\left[ \cos \left( \omega_{n} + \omega_{k} \right) \tau +
\cos \left( \omega_{n} - \omega_{k} \right) \tau \right] .
\end{displaymath}

An important consequence of Eq.~(\ref{avj}) is that, except in very special
cases (discussed in Sec.~\ref{longdis}), the exponential rate of energy
growth is the same {\em for all oscillators}. In fact, every action variable
increases exponentially in time with a rate given by the largest eigenvalue
of the ${\bf M}$ matrix.

\section{Longitudinal disorder}
\label{longdis}

As a first application of the previous results, we can consider the special
case in which the random potential depends only on the longitudinal
coordinate, i.e., $U(x,y) = U(x)$. We will refer to this case as
``longitudinal disorder''. This form of random potential has been considered
in a different context, i.e., that of many-mode waveguides with a rough
surface, where it has been christened as ``stratified disorder''~\cite{Izr04}.
Here we recover the results of that paper using our more general formalism.

In the special case in which $U$ depends only on the longitudinal coordinate
$x$, the matrix elements~(\ref{wmatrix}) take the simple form
\begin{equation}
W_{nk} = U(t) \delta_{nk} .
\label{wlongdis}
\end{equation}
Substituting this expression in Eqs.~(\ref{dmatrix1}) and~(\ref{dmatrix2}),
one obtains that the diffusion matrix becomes
\begin{displaymath}
D_{nk}(J) = 4 \lambda_{n} J_{n}^{2} \delta_{nk}
\end{displaymath}
where we have introduced the symbols
\begin{equation}
\lambda_{n} =
\frac{\varepsilon^{2}}{4\omega_{n}^{2}} \int_{0}^{\infty} d \tau \;
\langle U(t) U(t+\tau) \rangle \cos 2 \omega_{n} \tau .
\label{lyaplongdis}
\end{equation}
As a consequence, the Fokker-Planck equation~(\ref{fp1}) reduces to
\begin{displaymath}
\frac{\partial P}{\partial t}(J,t) = \sum_{n=0}^{N}
2 \lambda_{n} \frac{\partial}{\partial J_{n}} \left[ J_{n}^{2}
\frac{\partial P}{\partial J_{k}}(J,t) \right] .
\end{displaymath}
The solution of this Fokker-Planck equation corresponding to the initial
distribution
\begin{displaymath}
P(J,t=0) = \prod_{n=0}^{N} \delta \left( J_{n} - J_{n}(0) \right)
\end{displaymath}
is a product of log-normal distributions
\begin{equation}
P(J,t) = \prod_{n=0}^{N} \frac{1}{\sqrt{8 \pi \lambda_{n} t}}
\exp \left\{ - \frac{\left[ \log \left(J_{n}/J_{n}(0)\right) - 4\lambda_{n}t
\right]^{2}}{8 \lambda_{n} t} \right\} .
\label{lognorm}
\end{equation}

Using expression~(\ref{wlongdis}) one can also compute the matrix elements
of the $\bf M$ matrix which determines, via Eq.~(\ref{avj}), the time
evolution of the averaged action variables. In the case of longitudinal
disorder the $\bf M$ matrix takes the simple diagonal form
\begin{displaymath}
{\bf M}_{nk} = 4 \lambda_{n} \delta_{nk} .
\end{displaymath}
Both this result and the factorised distribution~(\ref{lognorm}) imply that,
in the case of longitudinal disorder, the parametric oscillators are
decoupled. The result could have been obtained {\em a priori} by noting
that if the random potential depends only on the longitudinal coordinate,
Fourier transforming the Schr\"{o}dinger equation~(\ref{schroe1})
gives a system of independent equations for the Fourier components.
Thus, the quasi-1D model~(\ref{schroe1}) is effectively decomposed into
$N+1$ strictly 1D systems, in agreement with the results of~\cite{Izr04}.

Besides being independent, the oscillators are also energetically unstable;
in fact, solving Eq.~(\ref{avj}) one obtains
\begin{displaymath}
\overline{J_{n}} = e^{4\lambda_{n} t} J_{n}(0) .
\end{displaymath}
The coefficients $4 \lambda_{n}$ are therefore the mean rates of exponential
growth of the energies of the oscillators (since the energy of the $n$-th
oscillator is proportional to the corresponding action variable, $E_{n}=
\omega_{n} J_{n}$). Taking into account that the rate of exponential increase
of the energy is four times the rate of exponential orbit
divergence~\cite{Tes01}, we are led to the conclusion that the
coefficients~(\ref{lyaplongdis}) represent the Lyapunov spectrum
of the dynamical system~(\ref{dynsys}) and, consequently, of the quasi-1D
model~(\ref{schroe1}).
The localisation length, therefore, is the equal to the inverse of the
minimum Lyapunov exponent. Which of the Lyapunov exponents~(\ref{lyaplongdis})
is the smallest depends on the specific form of the Fourier transform of the
binary correlator $\langle U(t)U(t+\tau) \rangle = \sigma^{2} \chi(\tau)$.

In the case of $\delta$-correlated disorder, the Fourier transform of
$\chi(\tau) = \delta(\tau)$, is simply the unity, $\tilde{\chi}(\omega) = 1$,
and the smallest Lyapunov exponent corresponds to the oscillator
with largest unperturbed frequency, i.e., the oscillator with $n = 0$.
Hence the inverse of the localisation length is
\begin{displaymath}
l^{-1} = \lambda_{0} = \frac{\varepsilon^{2}\sigma^{2}}{8 \omega_{0}^{2}}
= \frac{\varepsilon^{2} \sigma^{2}}{8 E} .
\end{displaymath}

When the binary correlation function is not a delta, however, the smallest
Lyapunov exponent is not determined only by the largest frequency, but also
by the behaviour of the power spectrum of the random potential
\begin{equation}
\tilde{\chi}(\omega) = \int_{-\infty}^{\infty} \chi(\tau)
\cos \omega \tau d \tau = \frac{1}{\sigma^{2}} \int_{-\infty}^{\infty}
\langle U(t+\tau) U(t) \rangle \cos \omega \tau d \tau.
\label{power}
\end{equation}
An important consequence is that the system can go through a delocalisation
transition if long-range correlations of the disorder make the Fourier
transform of two-point correlator~(\ref{power}) vanish in a specific frequency
interval. This phenomenon has already been analysed for 1D
systems~\cite{Izr99,Tes01,Tes02} and for quasi-1D waveguides with stratified
disorder~\cite{Izr04}.

The main point is that, given a Lyapunov exponent $\lambda_{n}(\omega)$ with
{\em any} specific dependence on the frequency $\omega$, it is possible
to find a random potential $U(t)$ that generates the pre-defined function
$\lambda_{n}(\omega)$.
Specifically, if the Lyapunov exponent $\lambda_{n}(\omega)$ is known,
then the power spectrum $\tilde{\chi}(\omega)=8 \omega^{2} \lambda_{n}(\omega)
/\sigma^{2}$ is also defined.
One can then determine a function $\beta(t)$ whose Fourier transform is
$\sqrt{\tilde{\chi}(\omega)}$ via the inversion formula
\begin{displaymath}
\beta(t) = \int_{-\infty}^{\infty} \sqrt{\tilde{\chi}(\omega)} e^{-i\omega t}
\frac{d \omega}{2 \pi} .
\end{displaymath}
A random potential which produces the desired behaviour of $\lambda_{n}
(\omega)$ is then obtained by taking the convolution of the function
$\beta(t)$ with a stochastic process $\eta(t)$ with zero mean and delta-shaped
correlation function. In other words, one can consider the potential
\begin{displaymath}
U(t) = \int_{-\infty}^{+\infty} ds \; \beta(s) \eta(t+s)
\end{displaymath}
where $\eta(t)$ is a white noise with
\begin{eqnarray*}
\langle \eta(t) \rangle = 0 & \mbox{ and} &
\langle \eta(t) \eta(t+\tau) \rangle = \delta(\tau) .
\end{eqnarray*}
Following this recipe, one can obtain for instance a random potential with
the long-ranged correlation function
\begin{equation}
\chi(\tau) = \frac{1}{\tau} \left( \sin \nu_{1} \tau -
\sin \nu_{2} \tau \right)
\label{chitau}
\end{equation}
which corresponds to the ``window'' power spectrum
\begin{displaymath}
\tilde{\chi} (\omega) = \left\{
\begin{array}{ll}
1       & \mbox{if } \nu_{1} < \omega < \nu_{2} \\
0       & \mbox{otherwise}
\end{array} \right. .
\end{displaymath}
In such a case, one has that the for every frequency $\omega_{n}$ that
falls outside of the interval $[\nu_{1}/2:\nu_{2}/2]$ the corresponding
Lyapunov exponent $\lambda_{n}$ vanishes (at least within the limits of the
second-order approximation considered here).
By shifting the frequencies $\nu_{1}$ and $\nu_{2}$, one can therefore
obtain a delocalisation transition as soon as the smallest Lyapunov
exponent vanishes; this corresponds to the electronic wavefunction having
one extended Fourier component. If more Lyapunov exponents vanish, the
number of extended Fourier components increases and the delocalisation
effect becomes more robust; in the extreme case when all Lyapunov exponents
vanish, the electronic wavefunction is not affected by the random potential.
If the phenomenon is considered from the point of view of the dynamical
system~(\ref{dynsys}) one has that for every vanishing Lyapunov exponent
there is an oscillator which becomes stable energy-wise. When all the
Laypunov exponent are zero, all the oscillators are stabilised, i.e., the
exponential divergence of the orbits is suppressed and the dynamics ceases
to be chaotic.
The possibility of making selected Lyapunov exponents vanish entails that
in solid-state models the transmission properties can exhibit anomalous and
unexpected features of selective transparency, as happens for waveguides.
We refer the reader to~\cite{Izr04} for a detailed discussion of this
phenomenon.

\section{Delocalisation transition in quasi-1D models}
\label{deloc}

In the previous section we have considered the special case in which the
random potential depends only on the longitudinal coordinate. The methods
of Sec.~\ref{fokpla} can be applied also to the a potential of the general
form $U = U(x,y)$, but in this case it is usually impossible to find the
analytic solution $P(J,t)$ of the Fokker-Planck equation~(\ref{fp1}) as well
as to solve with non-numerical methods the differential equation~(\ref{avj})
for the average action variables.
In fact, the dependency of the random potential on the transversal variable
produces coupling of the different oscillators, so that the coupling
matrix~(\ref{wmatrix}) is no longer diagonal. As a consequence, neither the
diffusion matrix $D_{nk}$ in the Fokker-Planck equation~(\ref{fp1}) nor the
evolution matrix $M_{nk}$ in Eq.~(\ref{avj}) are diagonal and this makes
exceedingly difficult to solve analytically both equations.

One can, however, use a different technique to determine the {\em sum} of
the positive Lyapunov exponents, or Kolmogorov entropy. Setting the latter
equal to zero then gives a sufficient condition for delocalisation of the
electronic states of the quasi-1D model~(\ref{schroe1}) and a sufficient
and necessary condition for the suppression of orbit instability in the
dynamical system~(\ref{dynsys}).
In the following subsection we define the Kolmogorov entropy and we show
how it can be computed in the second-order approximation. The reader
uninterested in technical details may skip this subsection and the
next, where a few special cases are analysed, and go to Subsec.~\ref{secdelo}
where delocalisation effects are discussed.

\subsection{The Kolmogorov entropy}

It is well known that in a Hamiltonian system with $\nu$ degrees of freedom
there are $2 \nu$ Lyapunov exponents which, due to the symplectic structure
of the dynamical equations, obey the symmetry relation
\begin{displaymath}
\lambda_{i} = - \lambda_{2 \nu - i +1}
\end{displaymath}
with $i =1,\ldots, \nu$ (see, for instance,~\cite{Lic92}).
Because of this relation, one has $\nu$ non-negative Lyapunov exponents.
For a deterministic Hamiltonian system, at least one of these exponents
vanishes; in the present case, however, the presence of a noisy term in
the Hamiltonian~(\ref{ham1}) ensures that, under normal circumstances, the
non-negative exponents are actually positive.
The Lyapunov exponents are defined as the exponential rate of local
divergence of initially nearby trajectories; this definition, however, can
be used operatively only to compute the largest Lyapunov exponent. Lesser
Lyapunov exponent can be determined using a technique devised by Benettin
et al.~\cite{Ben80}.
The main idea is that the sum of the largest $k$ exponents is equal to the
exponential rate of increase in time of the volume of a parallelepiped
spanned by $k$ independent vectors ( with $k \leq \nu$). The result does not
depend on the choice of the initial vectors.

In the present case, we are interested in the sum of all the positive
Lyapunov exponents; for this reason we consider the volume of the
parallelepiped spanned by $\nu = N+1$ linearly independent vectors
$\vec{\xi}^{(0)}(t), \ldots, \vec{\xi}^{(N)}(t)$.
Such a volume can be expressed as the square root of a Gram determinant
\begin{equation}
V^{(\nu)}(t) = \sqrt{|\det {\bf G}(t)|}
\label{volume}
\end{equation}
where ${\bf G}$ is the $(N+1) \times (N+1)$ matrix with elements
\begin{equation}
{\bf G}_{ij}(t) = \vec{\xi}^{(i)}(t) \cdot \vec{\xi}^{(j)}(t) .
\label{gram}
\end{equation}
The sum of the $N+1$ positive Lyapunov exponents can then be written
as
\begin{equation}
\sum_{i=0}^{N} \lambda_{i} = \lim_{T \rightarrow \infty} \Big\langle
\log \frac{V^{(\nu)}(T)}{V^{(\nu)}(0)} \Big\rangle .
\label{kolmoentro}
\end{equation}

To apply this prescription to the dynamical system~(\ref{dynsys}) it is
convenient to introduce the rescaled variables
\begin{displaymath}
\begin{array}{ccl}
x_{n} & = & \displaystyle \frac{p_{n}}{\sqrt{\omega_{n}}} \\
x_{N+1+n} & = & \displaystyle \sqrt{\omega_{n}} q_{n}
\end{array}
\end{displaymath}
with $n=0,\ldots,N$, so that the unperturbed motion of the $n$-th oscillator
reduces to a rotation in the $(x_{n},x_{N+1+n})$ plane.

In terms of the new variables, the dynamical equations~(\ref{dynsys})
take the form
\begin{equation}
\begin{array}{ccl}
\dot{x}_{n} & = & \displaystyle
- \omega_{n} x_{N+1+n} + \sum_{k=0}^{N} \frac{\varepsilon W_{nk}(t)}
{\sqrt{\omega_{n} \omega_{k}}} x_{N+1+k} \\
\dot{x}_{N+1+n} & = & \displaystyle
\omega_{n} x_{n} .
\end{array}
\label{rescaledeqs}
\end{equation}
The time evolution of the system can be expressed in terms of the
evolution operator ${\bf U}(t)$, defined by the relation
\begin{equation}
\left( \begin{array}{c} x_{0}(t) \\
                        \vdots \\
                        x_{2N+1}(t)
\end{array} \right) = {\bf U}(t)
\left( \begin{array}{c} x_{0}(0) \\
                        \vdots \\
                        x_{2N+1}(0)
       \end{array} \right)
\label{u}
\end{equation}
Let us introduce the $(2N+2) \times (2N+2)$ matrices ${\bf A}$ and ${\bf B}$,
defined in block form as
\begin{eqnarray*}
{\bf A} =
\left( \begin{array}{lr}
{\bf 0}      &  -{\bf \Omega} \\
{\bf \Omega} &   {\bf 0}
\end{array} \right)
& \mbox{ and} &
{\bf B}(t) =
\left( \begin{array}{cc}
{\bf 0}   &  \quad {\bf \Omega}^{-1/2} {\bf W}(t) {\bf \Omega}^{-1/2} \\
{\bf 0}   &   \quad {\bf 0}
\end{array} \right)
\end{eqnarray*}
where ${\bf W}(t)$ is the $(N+1) \times (N+1)$ matrix whose elements are
defined by Eq.~(\ref{wmatrix}) and ${\bf \Omega}$ is the $(N+1) \times (N+1)$
diagonal matrix
\begin{displaymath}
{\bf \Omega}_{nk} = \omega_{n} \delta_{nk} .
\end{displaymath}
Writing the dynamical equations~(\ref{rescaledeqs}) in matrix form and
taking into account Eq.~(\ref{u}), it is easy to see that the evolution
operator is the solution of the matrix differential equation
\begin{displaymath}
\dot{\bf U} = \left[ {\bf A} + \varepsilon {\bf B}(t) \right] {\bf U}
\end{displaymath}
with the initial condition ${\bf U}(0) = {\bf 1}$.
Going to the interaction representation, one can write the evolution
operator in the form
\begin{displaymath}
{\bf U}(t) = e^{{\bf A}t} {\bf U}_{I}(t)
\end{displaymath}
where the first factor is the unperturbed evolution operator,
\begin{displaymath}
e^{{\bf A}t} = \left( \begin{array}{cc}
\cos {\bf \Omega} t & -\sin {\bf \Omega} t \\
\sin {\bf \Omega} t & \cos {\bf \Omega} t \\
\end{array} \right)
\end{displaymath}
while the second factor obeys the equation
\begin{equation}
\dot{{\bf U}}_{I} = \varepsilon {\bf B}_{I}(t) {\bf U}_{I}
\label{compactdyneq}
\end{equation}
with
\begin{displaymath}
{\bf B}_{I}(t) = e^{-{\bf A}t} {\bf B}(t) e^{{\bf A}t} .
\end{displaymath}
For our purposes it is useful to write the evolution operator in block
form:
\begin{displaymath}
\begin{array}{lcr}
{\bf U} = \left( \begin{array}{cc}
{\bf U}^{(a)} & {\bf U}^{(c)} \\
{\bf U}^{(b)} & {\bf U}^{(d)} \\
\end{array} \right) & \mbox{ and}&
{\bf U}_{I} = \left( \begin{array}{cc}
{\bf U}_{I}^{(a)} & {\bf U}_{I}^{(c)} \\
{\bf U}_{I}^{(b)} & {\bf U}_{I}^{(d)} \\
\end{array} \right)
\end{array} .
\end{displaymath}
This decomposition allows one to obtain from Eq.~(\ref{compactdyneq}) the
dynamical equations for the left blocks of the evolution operator in
interaction representation
\begin{equation}
\begin{array}{ccl}
\dot{\bf U}_{I}^{(a)} & = & \varepsilon {\bf W}_{cs}(t) \; {\bf U}_{I}^{(a)}
+ \varepsilon {\bf W}_{cc}(t) \; {\bf U}_{I}^{(b)} \\
\dot{\bf U}_{I}^{(b)} & = & -\varepsilon {\bf W}_{ss}(t) \; {\bf U}_{I}^{(a)}
- \varepsilon {\bf W}_{sc}(t) \; {\bf U}_{I}^{(b)} \\
\end{array}
\label{blockdyneq}
\end{equation}
where we have introduced the new symbols
\begin{eqnarray*}
{\bf W}_{cc}(t) & = & \cos {\bf \Omega} t \; {\bf \Omega}^{-1/2} \;
{\bf W}(t) \; {\bf \Omega}^{-1/2} \; \cos {\bf \Omega}t \\
{\bf W}_{cs}(t) & = & \cos {\bf \Omega} t \; {\bf \Omega}^{-1/2} \;
{\bf W}(t) \;  {\bf \Omega}^{-1/2} \; \sin {\bf \Omega}t \\
{\bf W}_{sc}(t) & = & \sin {\bf \Omega} t \; {\bf \Omega}^{-1/2} \;
{\bf W}(t) \; {\bf \Omega}^{-1/2} \; \cos {\bf \Omega}t \\
{\bf W}_{ss}(t) & = & \sin {\bf \Omega} t \; {\bf \Omega}^{-1/2} \;
{\bf W}(t) \;  {\bf \Omega}^{-1/2} \; \sin {\bf \Omega}t \\
\end{eqnarray*}
The identity ${\bf U}_{I}(0) = {\bf 1}$ implies that Eqs.~(\ref{blockdyneq})
are to be solved with initial conditions
\begin{equation}
\begin{array}{lcr}
{\bf U}_{I}^{(a)}(0) = {\bf 1} & \mbox{ and} &
{\bf U}_{I}^{(b)}(0) = {\bf 0}.
\end{array}
\label{uinicon}
\end{equation}

We can now select as initial parallelepiped the $(N+1)$-dimensional cube
of edge $\Delta$ in the space of rescaled momenta. In other words, we
consider the set of initial vectors
\begin{displaymath}
\vec{\xi}^{(i)}_{k}(0) = \Delta \; \delta_{ik}
\end{displaymath}
where the index $i$ (which identifies the independent vectors) runs from
$0$ to $N$, while the index $k$ (which labels the components of each vector)
runs from $0$ to $2N+1$.
With this choice of the initial vectors, and remembering that the evolved
vectors can be written as $\vec{\xi}^{(i)}(t) = {\bf U}(t) \vec{\xi}^{(i)}(0)$,
one obtains that the matrix~(\ref{gram}) can be expressed in terms of the
left blocks of the evolution operator
\begin{displaymath}
{\bf G} = \left[ {{\bf U}^{(a)}}^{T} {\bf U}^{(a)} +
{{\bf U}^{(b)}}^{T} {\bf U}^{(b)} \right] \Delta^{2} =
\left[ {{\bf U}_{I}^{(a)}}^{T} {\bf U}_{I}^{(a)} +
{{\bf U}_{I}^{(b)}}^{T} {\bf U}_{I}^{(b)} \right] \Delta^{2}
\end{displaymath}
(where the symbol ${\bf M}^{T}$ denotes the transpose of the matrix $\bf M$).
Note that the matrix ${\bf G}$ has the same form in the interaction
representation and in the original representation.
Inserting this matrix in formula~(\ref{volume}), one obtains the volume of
the expanding parallelepiped; after substituting this result in
expression~(\ref{kolmoentro}) one arrives at the conclusion that the
sum of the positive Lyapunov exponents can be written as
\begin{eqnarray*}
\sum_{i=0}^{N} \lambda_{i} & = & \lim_{T \rightarrow \infty}
\frac{1}{2T} \Big\langle \log \det
\left[ {{\bf U}_{I}^{(a)}}^{T}(T) {\bf U}_{I}^{(a)}(T) +
{{\bf U}_{I}^{(b)}}^{T}(T) {\bf U}_{I}^{(b)}(T) \right] \Big\rangle \\
& = & \lim_{T \rightarrow \infty} \frac{1}{2T} \int_{0}^{T} dt \;
\Big\langle \frac{d}{dt} \log \det
\left[ {{\bf U}_{I}^{(a)}}^{T}(t) {\bf U}_{I}^{(a)}(t) +
{{\bf U}_{I}^{(b)}}^{T}(t) {\bf U}_{I}^{(b)}(t) \right] \Big\rangle .
\end{eqnarray*}
Using the fact that
\begin{displaymath}
\frac{d}{dt} \log \det {\bf M} =
{\rm{Tr}} \left( \dot{\bf M} {\bf M}^{-1} \right) ,
\end{displaymath}
one can express the Kolmogorov entropy in the form
\begin{equation}
\begin{array}{ccl}
\displaystyle
\sum_{i=0}^{N} \lambda_{i} & = & \displaystyle
\lim_{T \rightarrow \infty} \frac{1}{2T} \int_{0}^{T} dt \; \Big\langle
\varepsilon \; {\rm{Tr}} \left\{
\left[ {\bf W}_{cs}(t) + {\bf W}_{cs}^{T}(t) \right] {\bf Z}_{1}(t)
- \left[ {\bf W}_{sc}(t) + {\bf W}_{sc}^{T}(t) \right] {\bf Z}_{2}(t)
\right. \\
& + & \left.
\left[ {\bf W}_{cc}(t) - {\bf W}_{ss}^{T}(t) \right] {\bf Z}_{3}(t) +
\left[ {\bf W}_{cc}^{T}(t) - {\bf W}_{ss}(t) \right] {\bf Z}_{3}^{T}(t)
\right\} \Big\rangle
\end{array}
\label{kolentro}
\end{equation}
where we have introduced the operators
\begin{equation}
\begin{array}{ccc}
{\bf Z}_{1} & = & {\bf U}_{I}^{(a)}
\left[ {{\bf U}_{I}^{(a)}}^{T} {\bf U}_{I}^{(a)} +
{{\bf U}_{I}^{(b)}}^{T} {\bf U}_{I}^{(b)} \right]^{-1}
{{\bf U}_{I}^{(a)}}^{T} \\
{\bf Z}_{2} & = & {\bf U}_{I}^{(b)}
\left[ {{\bf U}_{I}^{(a)}}^{T} {\bf U}_{I}^{(a)} +
{{\bf U}_{I}^{(b)}}^{T} {\bf U}_{I}^{(b)} \right]^{-1}
{{\bf U}_{I}^{(b)}}^{T} \\
{\bf Z}_{3} & = & {\bf U}_{I}^{(b)}
\left[ {{\bf U}_{I}^{(a)}}^{T} {\bf U}_{I}^{(a)} +
{{\bf U}_{I}^{(b)}}^{T} {\bf U}_{I}^{(b)} \right]^{-1}
{{\bf U}_{I}^{(a)}}^{T} .\\
\end{array}
\label{zop}
\end{equation}

Taking into account the dynamical equations~(\ref{blockdyneq}) for the blocks
of the evolution operator, it is easy to see that the operators~(\ref{zop})
satisfy the differential equations
\begin{equation}
\begin{array}{ccl}
\dot{\bf Z}_{1} & = & \varepsilon \left[ {\bf W}_{cs} {\bf Z}_{1} +
{\bf Z}_{1} {\bf W}_{cs}^{T} + {\bf W}_{cc} {\bf Z}_{3} +
{\bf Z}_{3}^{T} {\bf W}_{cc}^{T} -
{\bf Z}_{1} \left( {\bf W}_{cs} + {\bf W}_{cs}^{T} \right) {\bf Z}_{1}
\right. \\
& + & \left.
{\bf Z}_{3}^{T} \left( {\bf W}_{sc} + {\bf W}_{sc}^{T} \right) {\bf Z}_{3}
- {\bf Z}_{1} \left( {\bf W}_{cc} - {\bf W}_{ss}^{T} \right) {\bf Z}_{3} -
{\bf Z}_{3}^{T} \left({\bf W}_{cc}^{T} -{\bf W}_{ss} \right) {\bf Z}_{1}
\right] \\
\dot{\bf Z}_{2} & = & \varepsilon \left[ - {\bf W}_{sc} {\bf Z}_{2} -
{\bf Z}_{2} {\bf W}_{sc}^{T} - {\bf W}_{ss} {\bf Z}_{3}^{T} -
{\bf Z}_{3} {\bf W}_{ss}^{T} + {\bf Z}_{2} \left( {\bf W}_{sc} +
{\bf W}_{sc}^{T} \right) {\bf Z}_{2}
\right. \\
& - & \left. {\bf Z}_{3} \left( {\bf W}_{cs} + {\bf W}_{cs}^{T} \right)
{\bf Z}_{3}^{T} - {\bf Z}_{3} \left( {\bf W}_{cc} - {\bf W}_{ss}^{T} \right)
{\bf Z}_{2} - {\bf Z}_{2} \left( {\bf W}_{cc}^{T} - {\bf W}_{ss} \right)
{\bf Z}_{3}^{T} \right] \\
\dot{\bf Z}_{3} & = & \varepsilon \left[ -{\bf W}_{ss} {\bf Z}_{1} -
{\bf W}_{sc} {\bf Z}_{3} + {\bf Z}_{3} {\bf W}_{cs}^{T} + {\bf Z}_{2}
{\bf W}_{cc}^{T} - {\bf Z}_{3} \left( {\bf W}_{cs} + {\bf W}_{cs}^{T} \right)
{\bf Z}_{1} \right. \\
& + & \left.
{\bf Z}_{2} \left( {\bf W}_{sc} + {\bf W}_{sc}^{T} \right) {\bf Z}_{3}
- {\bf Z}_{3} \left( {\bf W}_{cc} - {\bf W}_{ss}^{T} \right) {\bf Z}_{3} -
{\bf Z}_{2} \left( {\bf W}_{cc}^{T} - {\bf W}_{ss} \right) {\bf Z}_{1}
\right] \\
\end{array}
\label{zeq}
\end{equation}
These equations determine the time evolution of the operators~(\ref{zop})
together with the initial conditions
\begin{displaymath}
\begin{array}{ccc}
{\bf Z}_{1}(0) = {\bf 1}, & {\bf Z}_{2}(0) = {\bf 0}, & {\bf Z}_{3}(0) =
{\bf 0}
\end{array}
\end{displaymath}
which can be obtained by substituting in the definitions~(\ref{zop}) the
initial conditions~(\ref{uinicon}) for the left blocks of the evolution
operator.
The system of equations~(\ref{zeq}) can be solved perturbatively by
considering solutions of the form
\begin{displaymath}
{\bf Z}_{i}(t) = \sum_{n=0}^{\infty} \varepsilon^{n} {\bf Z}_{i}^{(n)}(t)
\end{displaymath}
Substituting these trial solutions into Eq.~(\ref{zeq}), one obtains
the simple result
\begin{displaymath}
\begin{array}{ccl}
{\bf Z}_{1}(t) & = & \displaystyle {\bf 1} + o\left( \varepsilon \right) \\
{\bf Z}_{2}(t) & = & \displaystyle o\left( \varepsilon \right) \\
{\bf Z}_{3}(t) & = & \displaystyle - \varepsilon \int_{0}^{t}
{\bf W}_{ss} (\tau) d \tau + o\left( \varepsilon \right) .
\end{array}
\end{displaymath}
Putting these expressions in Eq.~(\ref{kolentro}) one obtains that
the sum of the positive Lyapunov exponents is
\begin{equation}
\begin{array}{ccl}
\displaystyle \sum_{i=0}^{N} \lambda_{i} & = & \displaystyle
\sum_{n=0}^{N} \sum_{k=0}^{N}
\frac{\varepsilon^{2}}{8 \omega_{n} \omega_{k}} \int_{0}^{\infty}
\left\{ \left[ \langle W_{kn}(t) W_{kn}(t+\tau) \rangle +
\langle W_{nk}(t) W_{kn}(t+\tau) \rangle \right] \cos \left( \omega_{n}
+ \omega_{k} \right) \tau \right. \\
& + & \left. \displaystyle
\left[ \langle W_{kn}(t) W_{kn}(t+\tau) \rangle -
\langle W_{nk}(t) W_{kn}(t+\tau) \rangle \right] \cos \left( \omega_{n}
- \omega_{k} \right) \tau \right\}  d \tau + o(\varepsilon^{2})
\end{array}
\label{kolmogorov}
\end{equation}

This formula gives the sum of the positive Lyapunov exponents for the
dynamical system~(\ref{dynsys}). Note that the result has been derived
without considering any specific form of the coupling matrix $W_{nk}$.
Therefore it can be applied also to cases in which the coupling matrix
differs from the form~(\ref{wmatrix}). In particular, if one considers
a symmetric matrix, $W_{nk} = W_{kn}$, the second term in the r.h.s. of
Eq.~(\ref{kolmogorov}) vanishes and the result reduces to the form obtained
in~\cite{Han89} for a similar problem. Here, however, the $W_{nk}$ matrix
is not fully symmetric, because of the Kronecker delta in Eq.~(\ref{wmatrix})
which can be traced back to the fact that the zero-th Fourier mode has no
twin component unlike the other modes (which come in equal pairs $\tilde{U}_{n}
= \tilde{U}_{-n}$). However, the second term in Eq.~(\ref{kolmogorov})
can be neglected in the limit of a large number of Fourier components, in
which case the asymmetry linked to the zero-th channel becomes
negligible.

To analyse the physical implications of Eq.~(\ref{kolmogorov}), it is
useful to express the Kolmogorov entropy in terms of the Fourier components
of the random potential $U$. Substituting the matrix elements~(\ref{wmatrix})
in Eq.~(\ref{kolmogorov}), one obtains
\begin{equation}
\begin{array}{l}
\displaystyle
\sum_{i=0}^{N} \lambda_{i} = \sum_{n=1}^{N}
\frac{1}{8 \omega_{n} \omega_{0}} \int_{0}^{\infty}
\langle \tilde{U}_{n}(t) \tilde{U}_{n}(t+\tau) \rangle
\left[ \cos \left( \omega_{n} + \omega_{0} \right) \tau +
\cos \left( \omega_{n} - \omega_{0} \right) \tau \right] \; d\tau \\
\displaystyle
+ \sum_{n=-N}^{N} \sum_{k=-N}^{N}
\frac{1}{8 \omega_{n} \omega_{k}} \int_{0}^{\infty}
\left[ \langle \tilde{U}_{n+k}(t) \tilde{U}_{n+k}(t+\tau) \rangle +
\langle \tilde{U}_{n-k}(t) \tilde{U}_{n+k}(t+\tau) \rangle \right]
\cos \left( \omega_{n} + \omega_{k} \right) \tau  \; d \tau .\\
\end{array}
\label{almostfinal}
\end{equation}
Note that we have set the bookkeeping parameter $\varepsilon =1$, as we
will do from now on, with the tacit understanding that all results are
valid within the second-order approximation.
Eq.~(\ref{almostfinal}) represents the Kolmogorov entropy for any kind
of random potential.
We observe that the second sum on the r.h.s. of Eq.~(\ref{almostfinal})
contains a number of terms of order $O(N^{2})$ and therefore for $N \gg 1$
it is dominant with respect to the first sum which has only $O(N)$ terms.
Hence if the number of oscillators/modes $N$ is large, one can approximate
the Kolmogorov entropy~(\ref{almostfinal}) with
\begin{equation}
\begin{array}{ccl}
\displaystyle
\sum_{i=0}^{N} \lambda_{i} & \simeq & \displaystyle
\sum_{n=-N}^{N} \sum_{k=-N}^{N}
\frac{1}{8 \omega_{n} \omega_{k}} \int_{0}^{\infty}
\left[ \langle \tilde{U}_{n+k}(t) \tilde{U}_{n+k}(t+\tau) \rangle \right. \\
& + & \displaystyle \left.
\langle \tilde{U}_{n-k}(t) \tilde{U}_{n+k}(t+\tau) \rangle \right]
\cos \left( \omega_{n} + \omega_{k} \right) \tau  \; d \tau .
\end{array}
\label{result1}
\end{equation}
We remember that the binary correlators which appear in Eq.~(\ref{result1})
are the correlators of the Fourier components~(\ref{uft}) of the random
potential.
They are linked to the binary correlator~(\ref{bincor}) via the double
Fourier transform
\begin{equation}
\langle \tilde{U}_{n}(t) \tilde{U}_{k}(t+\tau) \rangle =
\frac{\sigma^{2}}{4L^{2}} \int_{-L}^{L} dy \int_{-L}^{L} dy' \;
\chi(\tau, y-y') \cos \frac{\pi n y}{L} \cos \frac{\pi k y'}{L}.
\label{corruft}
\end{equation}
If the correlators~(\ref{corruft}) decay sufficiently fast as a function
of the difference $|n-k|$ of the indices of the Fourier components of the
potential, the second correlator in Eq.~(\ref{result1}) gives a marginal
contribution with respect to the first one. Therefore one can replace
expression~(\ref{result1}) with
\begin{equation}
\sum_{i=0}^{N} \lambda_{i} \simeq
\sum_{n=-N}^{N} \sum_{k=-N}^{N}
\frac{1}{8 \omega_{n} \omega_{k}} \int_{0}^{\infty}
\langle \tilde{U}_{n+k}(t) \tilde{U}_{n+k}(t+\tau) \rangle 
\cos \left( \omega_{n} + \omega_{k} \right) \tau  \; d \tau .
\label{result2}
\end{equation}

\subsection{Application to specific cases}

We will now apply the general formulae derived in the previous subsection
to a few particular cases and see how one can recover known specific results
from the general expressions~(\ref{almostfinal}) and~(\ref{result2}).
Let us consider first the case of longitudinal disorder, i.e., of a random
potential of the form $U(x,y) = U(x)$.
In this case the only non-vanishing Fourier component~(\ref{uft}) is
the zero-th one
\begin{displaymath}
\tilde{U}_{n}(t) = \delta_{n0} U(t)
\end{displaymath}
and the binary correlators~(\ref{corruft}) become
\begin{equation}
\langle \tilde{U}_{n}(t) \tilde{U}_{k}(t+\tau) \rangle =
\sigma^{2} \chi(\tau) \delta_{n0} \delta_{k0} .
\label{ldcorruft}
\end{equation}
These correlators vanish unless $n=k=0$ and therefore there is no doubt that
they decay fast for increasing values of $|n-k|$. Hence the approximate
formula~(\ref{result2}) can be applied; inserting expression~(\ref{ldcorruft})
in Eq.~(\ref{result2}) one obtains
\begin{equation}
\sum_{n=0}^{N} \lambda_{n} \simeq \sum_{n=0}^{N} \frac{\sigma^{2}}{4
\omega_{n}^{2}} \int_{0}^{\infty} \chi(\tau) \cos 2 \omega_{n} \tau \;
d \tau \left( 1 - \frac{1}{2} \delta_{n0} \right) .
\label{apprldkolmo}
\end{equation}

The exact result can be obtained by substituting the
correlators~(\ref{ldcorruft}) in Eq.~(\ref{almostfinal}); this gives
\begin{equation}
\sum_{n=0}^{N} \lambda_{n} = \sum_{n=0}^{N} \frac{\sigma^{2}}{4
\omega_{n}^{2}} \int_{0}^{\infty} \chi(\tau) \cos 2 \omega_{n} \tau \;
d \tau
\label{ldkolmo}
\end{equation}
and by comparing Eq.~(\ref{apprldkolmo}) with Eq.~(\ref{ldkolmo}) we see
that the difference between the two expressions is indeed negligible in
the large $N$ limit.
We remark that Eq.~(\ref{ldkolmo}) is in perfect agreement with what one
obtains for the sum of the positive Lyapunov exponents~(\ref{lyaplongdis})
derived in Sec.~\ref{longdis}. From the physical point of view,
Eq.~(\ref{ldkolmo}) is a natural consequence of the fact that, as discussed
in  Sec.~\ref{longdis}, in the case of longitudinal disorder the Fourier
modes/oscillators are decoupled so that the quasi-1D model reduces to a
sum of strictly 1D systems.

As a special application of formula~(\ref{ldkolmo}) one can consider the
case of longitudinal white noise. In this case $\chi(\tau) = \delta(\tau)$
and the Kolmogorov entropy~(\ref{ldkolmo}) takes the form
\begin{displaymath}
\sum_{n=0}^{N} \lambda_{n} = \sum_{n=0}^{N} \frac{\sigma^{2}}{8
\omega_{n}^{2}}
\end{displaymath}
in agreement with the well-known expression
\begin{displaymath}
\lambda_{n} = \frac{\sigma^{2}}{8 \omega_{n}^{2}}
\end{displaymath}
for the Lyapunov exponent of a noisy 1D oscillator of frequency $\omega_{n}$.

Eq.~(\ref{ldkolmo}) can be used also to derive the inverse localisation
length for a 1D model, which can be seen as the limit case of a quasi-1D
model when the transversal dimensions go to zero, $L \rightarrow 0$.
In this limit the only elliptic Fourier component is the zero-th one and,
indeed, setting $N=0$ in Eq.~(\ref{ldkolmo}) one recovers the well-known
expression~(\ref{invloc1d}) for the inverse localisation length in 1D models.

Let us now consider the case of a random potential which is a white noise in
both the longitudinal and the transversal direction. In this case the binary
correlator~(\ref{bincor}) has the form
\begin{displaymath}
\chi(x,y) = \delta(x) \delta(y)
\end{displaymath}
and the correlators of the Fourier components~(\ref{uft}) are
\begin{equation}
\langle \tilde{U}_{n}(t) \tilde{U}_{k}(t + \tau) \rangle =
\frac{\sigma^{2}}{4L} \delta(\tau)
\left( \delta_{n,k} + \delta_{n,-k} \right) .
\label{wncorruft}
\end{equation}
The form of the correlators~(\ref{wncorruft}) implies that the second
correlator in the expression~(\ref{result1}) gives only a marginal
contribution, so we can substitute Eq.~(\ref{wncorruft}) in
formula~(\ref{result2}).
Neglecting terms with $O(N)$ addends, one obtains
\begin{equation}
\sum_{i=0}^{N} \lambda_{i} \simeq
\frac{\sigma^{2}}{16 L} \left( \sum_{n=0}^{N}
\frac{1}{\omega_{n}} \right)^{2} .
\label{kolmown}
\end{equation}
This equation coincides with the result obtained for a discrete model by
Hansel and Luciani in~\cite{Han87} if one identifies the width $2L$ of the
doubled strip and the frequencies $\omega_{n}$ in~(\ref{kolmown})
respectively with the number of channels and the square roots of the energies
in the model of Hansel and Luciani.

\subsection{A sufficient condition for delocalisation}
\label{secdelo}

The importance of Eq.~(\ref{almostfinal}) lies in the fact that it
provides a criterion for the onset of a delocalisation transition in
quasi-1D models. In fact, the condition
\begin{equation}
\sum_{i=0}^{N} \lambda_{i} = 0
\label{delocond}
\end{equation}
represents both a necessary and sufficient condition for the suppression
of orbit instability in the system of parametric oscillators~(\ref{dynsys})
and a sufficient condition for the localisation length to diverge. That
condition~(\ref{delocond}) is not necessary for the delocalisation of the
electronic states depends on the fact that, as discussed in Sec.~\ref{analogy},
the localisation length is equal to the {\em smallest} Lyapunov exponent.
Therefore, delocalisation sets in as soon as the minimum Lyapunov exponent
vanishes, even if larger Lyapunov exponents are non-zero.
On the contrary, in the case of the dynamical system~(\ref{dynsys}), unless
{\em all} Lyapunov exponents vanish, initially nearby orbits exponentially
diverge.

The importance of expression~(\ref{almostfinal}) rests on the fact that
it allows one to prove that, for specific kinds of long-range correlated
disorder, the condition~(\ref{delocond}) is fulfilled over a certain range
of the electronic energy and, therefore, a continuum of extended states
arises even in quasi-1D models. The effect is analogous to the one observed
in strictly 1D models~\cite{Izr99, IDKT04}.
To see how a delocalisation transition can occur, one can observe that
the Kolmogorov entropy~(\ref{almostfinal}) vanishes if the Fourier transforms
in the longitudinal direction of the binary correlators~(\ref{corruft})
are zero
\begin{equation}
\int_{0}^{\infty}
\langle \tilde{U}_{n_{1}}(t) \tilde{U}_{n_{2}}(t+\tau) \rangle
\cos \omega \tau \; d \tau = 0
\label{deloc1}
\end{equation}
for all values of the indices $n_{1}$ and $n_{2}$ and for every frequency
$\omega$ in the interval $[0:2\sqrt{E}]$. The frequency interval is determined
by taking into account that the frequencies~(\ref{freque}) vary in the interval
$[0:\sqrt{E}]$ and that formula~(\ref{almostfinal}) contains cosines with
frequency up to twice the maximum value of the frequencies $\omega_{n}$.
Condition~(\ref{deloc1}), however, is equivalent to the requirement that
the Fourier transform of the correlator~(\ref{bincor})
\begin{equation}
\tilde{\chi}(\omega_{x}, \omega_{y}) =
\int_{0}^{\infty} dx \int_{0}^{\infty} dy \; \chi(x,y)
\cos \omega_{x} x \cos \omega_{y} y
\label{bincor2}
\end{equation}
should vanish for $0 \leq \omega_{x} \leq 2\sqrt{E}$, i.e.,
\begin{eqnarray}
\tilde{\chi}(\omega_{x},\omega_{y}) = 0 & \mbox{ for} &
0 \leq \omega_{x} \leq 2\sqrt{E} .
\label{deloc2}
\end{eqnarray}
A potential whose binary correlator satisfies this condition can be
constructed with a slight generalisation of the method discussed in
Sec.~\ref{longdis} for the longitudinal disorder case.
Starting from a binary correlator~(\ref{bincor2}) with arbitrary
frequency dependence, one can obtain the function
\begin{displaymath}
\beta(x,y) = \int_{-\infty}^{\infty} \frac{d \omega_{x}}{2 \pi}
\int_{-\infty}^{\infty} \frac{d \omega_{y}}{2 \pi}
\sqrt{\tilde{\chi}(\omega_{x},\omega_{y})}
\exp \left(-i\omega_{x} x - i \omega_{y} y \right) .
\end{displaymath}
The required potential can then be constructed via the convolution
product
\begin{displaymath}
U(x,y) = \sigma^{2}
\int_{-\infty}^{\infty} ds_{x} \int_{-\infty}^{\infty} ds_{y} \;
\beta(s_{x},s_{y}) \eta(s_{x}+x, s_{y}+y)
\end{displaymath}
where $\eta$ is a stochastic process with
\begin{eqnarray*}
\langle \eta(x,y) \rangle = 0 & \mbox{ and} &
\langle \eta(x,y) \eta(x',y') \rangle = \delta (x-x') \delta (y-y') .
\end{eqnarray*}

Following this procedure, one can obtain a potential which fulfils the
delocalisation condition~(\ref{deloc2}). For example, one can consider
a potential with long-range correlations in the longitudinal direction
of the form
\begin{equation}
\chi(x,y) = \frac{1}{x} \left( \sin \nu_{1} x - \sin \nu_{2} x \right)
\gamma(y)
\label{chitau2}
\end{equation}
where $\gamma(y)$ is the correlation function in the transversal direction.
The binary correlator~(\ref{chitau2}) is the simplest possible generalisation
of the correlator~(\ref{chitau}). Its power spectrum has the form
\begin{displaymath}
\tilde{\chi} (\omega_{x},\omega_{y}) = \left\{
\begin{array}{ll}
\tilde{\gamma}(\omega_{y}) & \mbox{if } \nu_{1} < \omega_{x} < \nu_{2} \\
0                          & \mbox{otherwise}
\end{array} \right. .
\end{displaymath}
By selecting a potential such that $\nu_{1}=2\sqrt{E}$, one is ensured
that the delocalisation condition~(\ref{delocond}) is fulfilled.
We are thus led to the conclusion that a delocalisation transition can occur
if the disorder exhibits long-range correlations that make the power spectrum
vanish in an appropriate frequency interval.

Ons should hasten to add that this conclusion is rigorously true only
for weak disorder and within the second-order approximation. In 1D models
it has been shown that the delocalisation transition produced by long-range
correlations of the disorder is a second-order effect and the ``extended
states'' are, in fact, electronic states which extend over a spatial
range of order $O(1/\sigma^{4})$ rather than
$O(1/\sigma^{2})$~\cite{Tes02}.
The same result can be expected for quasi-1D models; this does not diminish
the practical importance of the delocalisation transition analysed here
because, for weak disorder, i.e., when $\sigma^{2} \rightarrow 0$, the
increase of the spatial range of the electronic wavefunction can be huge,
being of order $O(1/\sigma^{2})$. For finite samples, therefore, the
delocalisation can be real and manifest itself in a strong change of the
transport properties of the disordered sample.

\section{Conclusions}
\label{conclu}

In this work we have shown how the spatial structure of electronic states
in a quantum quasi-1D model with weak disorder can be analysed in term of
the time evolution of a classical system of parametric oscillators with
weak stochastic couplings.
By Fourier-transforming the stationary Schr\"{o}dinger equation for the
quasi-1D model in the transversal directions, one obtains a set of
equations for the Fourier components of the electronic wavefunction
that can be mapped unto the dynamical equations of a Hamiltonian
system of coupled oscillators.
The spatial behaviour of the Fourier components of the wavefunction is thus
matched to the time evolution of the oscillators, while the disorder in the
quasi-1D model manifests itself as noise in its dynamical analogue.
The specific effect of the noise is to perturb the frequencies of the
oscillators and to couple the oscillators among themselves.
Both models are characterised by a set of characteristic Lyapunov exponents;
however, whereas in the solid state model the key Lyapunov exponent is the
smallest one, which is equal to the inverse of the localisation length, in
the dynamical system the most important exponent is the largest, which
defines the mean rate of exponential divergence of the orbits.

The dynamics of the Hamiltonian system can be analysed in full detail when
the random potential in the quasi-1D model depends only on the longitudinal
coordinate. In this case the oscillators are effectively decoupled and it
is possible to obtain the whole Lyapunov spectrum of the system.
One can thus make a complete study of the delocalisation effects produced
by specific long-range correlation of the disorder.

The general case of a random potential which depends both on longitudinal
and transversal coordinates is more difficult to handle; nevertheless, in
it is possible to evaluate perturbatively the sum of the positive Lyapunov
exponents (or Kolmogorov entropy).
Using this result, one can show that specific kinds of long-range correlated
disorder make all Lyapunov exponents vanish within the second-order
approximation and therefore produce a delocalisation transition in quasi-1D
models like they do in strictly 1D models.

\section{Acknowledgements}

L.T. acknowledges the support of Proyecto 4.17 of the CIC of the Universidad
Michoacana de San Nicol\'{a}s de Hidalgo (Mexico). F.M.I. acknowledges
the support by the CONACYT (Mexico) grant No.~43730.

\end{document}